\documentclass[printer]{aa} 
\usepackage{amsmath}
\usepackage{color}
\usepackage{graphicx}
\usepackage{txfonts}
\usepackage{natbib,twoopt}
\bibpunct{(}{)}{;}{a}{}{,}
\usepackage[breaklinks=true]{hyperref} 
\bibpunct{(}{)}{;}{a}{}{,} 
\makeatletter
\newcommandtwoopt{\citeads}[3][][]{\href{http://adsabs.harvard.edu/abs/#3}%
{\def\hyper@linkstart##1##2{}%
\let\hyper@linkend\@empty\citealp[#1][#2]{#3}}}
\newcommandtwoopt{\citepads}[3][][]{\href{http://adsabs.harvard.edu/abs/#3}%
{\def\hyper@linkstart##1##2{}%
\let\hyper@linkend\@empty\citep[#1][#2]{#3}}}
\newcommandtwoopt{\citetads}[3][][]{\href{http://adsabs.harvard.edu/abs/#3}%
{\def\hyper@linkstart##1##2{}%
\let\hyper@linkend\@empty\citet[#1][#2]{#3}}}
\newcommandtwoopt{\citeyearads}[3][][]%
{\href{http://adsabs.harvard.edu/abs/#3}
{\def\hyper@linkstart##1##2{}%
\let\hyper@linkend\@empty\citeyear[#1][#2]{#3}}}
\makeatother

\begin{document}


\title{Eclipses of the inner satellites of Jupiter observed in 2015\thanks{Three of the four eclipse observations where recorded at the 1 m telescope of Pic du Midi Observatory (S2P), the other at Saint-Sulpice Observatory.}}
\author{E. Saquet \inst{1,2}\and N. Emelyanov\inst{3,2}\and F. Colas\inst{2}\and J.-E. Arlot\inst{2}\and  V. Robert\inst{1,2}\and B. Christophe\inst{4}\and O. Dechambre\inst{4} }
\institute{
Institut Polytechnique des Sciences Avanc\'ees IPSA, 11-15 rue Maurice Grandcoing, 94200 Ivry-sur-Seine, France
\and
IMCCE, Observatoire de Paris, PSL Research University, CNRS-UMR 8028, Sorbonne Universités, UPMC, Univ. Lille 1, 77 Av. Denfert-Rochereau, 75014 Paris, France.
\and
M. V. Lomonosov Moscow State University – Sternberg astronomical institute, 13 Universitetskij prospect, 119992 Moscow, Russia 
\and
Saint-Sulpice Observatory, Club Eclipse, Thierry Midavaine, 102 rue de Vaugirard, 75006 Paris, France\\
\email{eleonore.saquet@obspm.fr; eleonore.saquet@ipsa.fr}
}
\date{Received xxxx xx, 2016; accepted xxxx xx, 2016}


\abstract
{}
{During the 2014-2015 campaign of mutual events, we recorded   ground-based photometric observations of eclipses of  \object{Amalthea} (JV) and, for the first time, \object{Thebe} (JXIV)  by the Galilean moons. We focused on estimating whether the positioning accuracy of the inner satellites determined with photometry is sufficient for dynamical studies.}   
 {We observed two eclipses of Amalthea and one of Thebe with the 1 m  telescope at Pic du Midi Observatory using an IR filter and a mask placed over the planetary image to avoid blooming features. A third observation of Amalthea was taken at Saint-Sulpice Observatory with a 60 cm telescope using a methane filter (890 nm) and a deep absorption band to decrease the contrast between the planet and the satellites. After background removal, we computed a differential aperture photometry to obtain the light flux, and followed with an astrometric reduction.}   
 {We provide astrometric results with an external precision of 53 mas for the eclipse of Thebe, and 20 mas for that of Amalthea. These observation accuracies largely override standard astrometric measurements. The $(O-C)$s for the eclipse of Thebe are $75$ mas on the $X$-axis and $120$ mas on the $Y$-axis. The $(O-C)$s for the total eclipses of Amalthea are $95$ mas and 22 mas, along the orbit, for two of the three events. Taking into account the ratio of $(O-C)$ to precision of the astrometric results, we show a significant discrepancy with the theory established by Avdyushev and Ban'shikova in 2008, and the JPL JUP 310 ephemeris.}
  {}
\authorrunning{E. Saquet et al.}
\maketitle


\section{Introduction}

The regular satellite system of \object{Jupiter} is composed of two groups: the Galilean moons and the inner satellites. Metis, Adrastea, Amalthea, and Thebe  orbit close to Jupiter. The mean radii from the planet are 1.79 Jupiter radius ($R_j$), $1.81\,R_j$, $2.54\,R_j$, and  $3.11\,R_j$, respectively. Amalthea, the biggest and brightest inner satellite, was discovered by \citetads{1892AJ.....12...81B} with the 36-inch refractor at Lick Observatory in California, USA. The three other inner moons were discovered by Voyager 1. Their interaction with the ring of Jupiter \citepads{1999Sci...284.1146B} and the Galilean satellites has made them interesting to study, in particular  for dynamic purposes.

The proximity of the inner satellites to Jupiter make them difficult targets for ground-based observations, resulting in a lack of precision in their respective dynamical models compared to the Galilean moons. As a consequence,  observers rely on different techniques such as a methane absorption band filter, coronagraphs, or infra-red cameras to be able to observe these objects.

Direct astrometry could be used to improve the ephemerides, but it encounters difficulties in detecting and using reference stars because the brightness of Jupiter produces an important background photon noise \citepads{colas_1991}. On the other hand and in the case of the inner satellites, mutual phenomena observation is the best way to make accurate astrometry since we only need to observe one object. Mutual events have been observed since the 1970s for the Galilean moons and have contributed to improving considerably their dynamical models. The inner moons were   observed during the previous mutual event season \citepads{2010A&A...522A...6C}.

From August 2014 to August 2015 a mutual phenomena period occurred and we recorded four of them: three eclipses of Amalthea, and the first one with Thebe ever observed from the ground. We focused our observations on estimating whether the positioning accuracy of the inner satellites determined with photometry is sufficient for dynamical studies.


\section{Mutual events}

\begin{figure}
\centering
\includegraphics[width=\hsize]{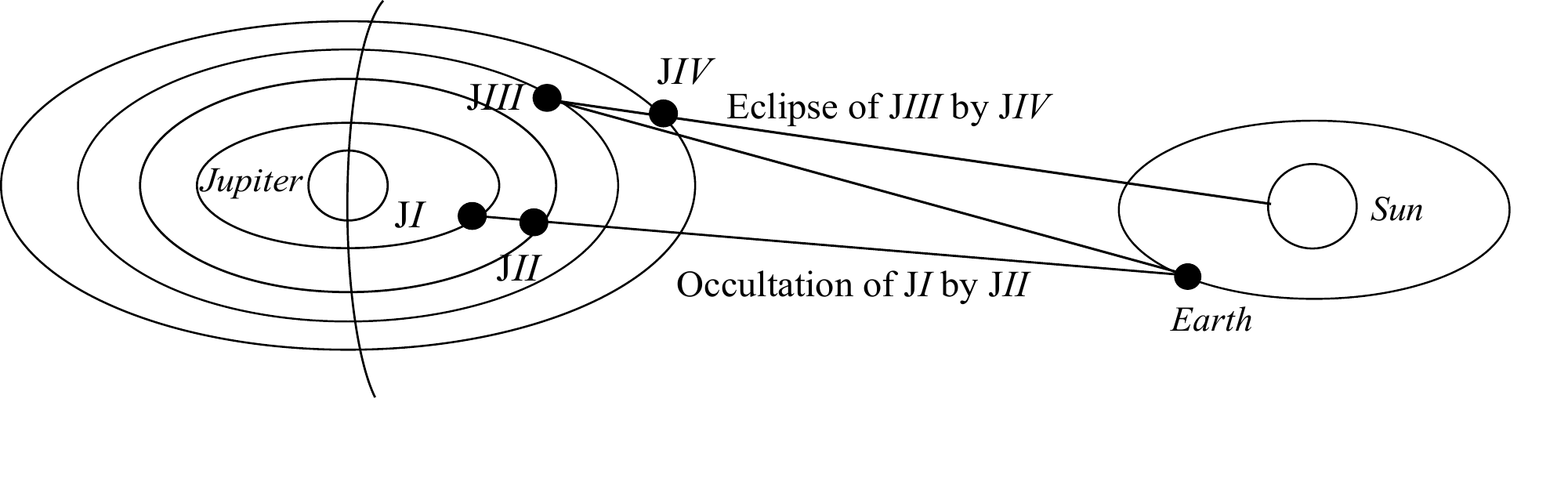}
\caption{\normalsize{Principle of an occultation and an eclipse of the satellites of Jupiter.}}
\label{FigPhemu}
\end{figure}

When the common Galilean orbital plane crosses the ecliptic plane, mutual events can occur (Fig. \ref{FigPhemu}). Depending on the configuration from the Earth or the Sun, we will speak about occultations or eclipses. Occultations can occur when the Joviocentric declination of the Earth becomes zero, eclipses can occur when the Joviocentric declination of the Sun becomes zero.

Mutual events are more accurate than the other observation methods because effects due to the atmosphere that could affect observational data are eliminated. Furthermore, the recording of a timing is easier than in astrometric measurement \citepads{2014A&A...572A.120A}. However, such rare and accurate observations imply numerous efforts to make as many observations as possible during the short favorable periods. Because the visibility of a phenomenon depends on the location of the observer,  international cooperation is necessary to cover the maximum range of geographic longitudes, as  has already been discussed in \citetads{2014A&A...572A.120A} and in \citetads{2007ASPC..370...58A}. Photometric records of mutual events are useful to improve the satellite ephemerides mainly because the determined positioning accuracy is more precise than those derived from direct astrometry. The precision of the mutual event observations is up to 20 mas for the Galilean moons \citepads{2014A&A...572A.120A}, or 60 km at the distance of Jupiter. Thus, we decided to adapt the methods for the inner moons.

During the 2014-2015 period, a series of occultations and eclipses occurred among the satellites of Jupiter thanks to the equinox on the planet in February 2015. We recorded four eclipses of the inner satellites: one of Amalthea by \object{Callisto} (4E5) and one by \object{Ganymede} (3E5), and one of Thebe by Callisto (4E14) at Pic du Midi Observatory, and a last one of Amalthea by \object{Europa} (2E5) at Saint-Sulpice Observatory.


\section{Observations and data reductions}

The predictions of the observed eclipses are given in Table \ref{table:1}, the observational settings in Table \ref{table:2}.

\begin{table*}
\caption{\normalsize{Predictions of the observed mutual events.}}            
\label{table:1}      
\centering                          
\begin{tabular}{c c c c c c c c c}        
\hline\hline                 
Date & Observation & UT start      & UT end       &Type & Dur. &Impact   &Magnitude\\    
        &    site           &(hh:mm:ss) & (hh:mm:ss) & & (m)          &              &\\
\hline                        
\textbf{2015 Jan 7}& T1M - Pic du Midi & 04:45:40 & 04:53:07 &   4E14  &   7.5&  0.234  & 16.2\\
\textbf{2015 Jan 7} & T1M - Pic du Midi & 05:23:56  & 06:15:20 &   4E5  &  51.4 & 0.999 &  14.3\\
\textbf{2015 Apr 7} & T1M - Pic du Midi & 19:51:45  & 19:53:59  &  3E5   &  2.2 & 0.837   &14.5\\
\textbf{2015 Apr 8} & T60 - Saint-Sulpice & 20:16:17  &   20:18:59 &   2E5 &    2.7&  0.190 &  14.5\\
\hline                                   
\end{tabular}
\end{table*}

\begin{table*}
\caption{\normalsize{Observations setting of the observed mutual events.}}             
\label{table:2}      
\centering                          
\begin{tabular}{c c c c c c c c}        
\hline\hline                 
Date &Type & Observation & Exp. & Filter & UT of         & UT of       & Number \\    
        &         &   site            & Time &          & first frame & last frame & of frame\\
        &         &                     & (s)    &          & (hh:mm:ss)&(hh:mm:ss)& \\
\hline                        
\textbf{2015 Jan 7}&   4E14 & T1M - Pic du Midi  & 20 & $Rg_{mask}$ & 04:34:43 & 04:58:12   &   62\\
\textbf{2015 Jan 7} &  4E5 & T1M - Pic du Midi & 20 & $Rg_{mask}$  & 05:05:48  & 06:29:31 &  206\\
\textbf{2015 Apr 7} &  3E5 & T1M - Pic du Midi & 15 & $Rg_{mask}$  & 19:49:56  & 20:03:38 & 49\\
\textbf{2015 Apr 8} &  2E5 & T60 - Saint-Sulpice & 5 & 890 nm & 20:05:11  &   20:35:06  & 360\\
\hline                                   
\end{tabular}
\end{table*}

\subsection{Observations at Pic du Midi Observatory}

We used the 1 m telescope of the Pic du Midi Observatory  (IAU Observatory Code: 586) to record the events of 4E14 on 2015 Jan 7, 4E5 on 2015 Jan 7, and 3E5 on 2015 Apr 7. The camera was an Andor Ikon-L. To reduce the blooming effect of the saturated image of Jupiter, we used a Schott RG695 lowpass filter with a mask placed over the planetary image. To ensure a timing accuracy of 1 ms, the camera shutter was triggered with a GPS card. The pixel size was of 0.4 arcsec on the sky.

\begin{figure}
\centering
\resizebox{\hsize}{!}{\includegraphics{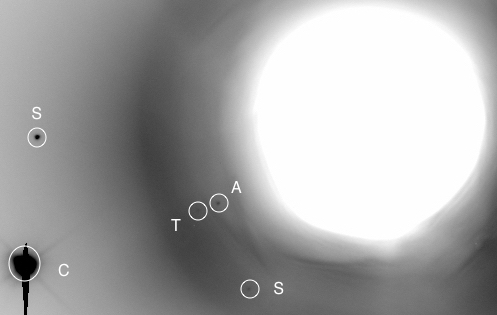}}
\caption{{\normalsize Subframe of a typical image (negative) during the observation of the 4E5 event on 2015 Jan 7 at Pic du Midi Observatory, France. The large white area is the mask used to reduce the intensity of Jupiter. We observe Amalthea (A), Thebe (T), two stars (S), and the bright source saturating the detector Callisto (C).}}
\label{FigPic}
\end{figure}

\begin{figure}
\centering
\resizebox{\hsize}{!}{\includegraphics{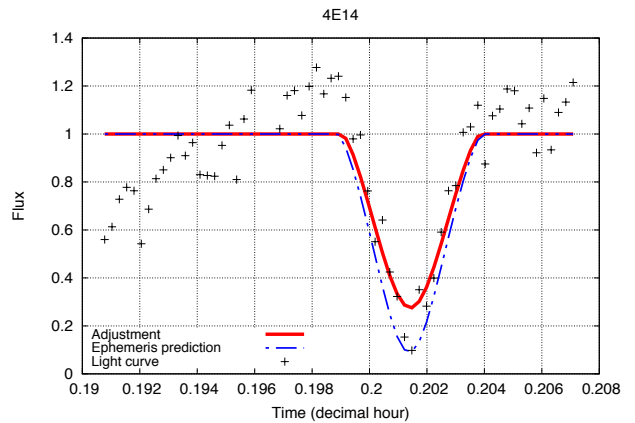}}
\caption{{\normalsize Fitted light curve (flux versus time in decimal hour) of the 4E14  event. Callisto eclipses Thebe on 2015 Jan 7. The blue dot-dash line denotes the theoretical flux drop according to  JUP310 ephemeris \citepads{JUP310}, the red line denotes the fitted flux drop of the event.}}
\label{Fig4E6}
\end{figure}

\begin{figure}
\centering
\resizebox{\hsize}{!}{\includegraphics{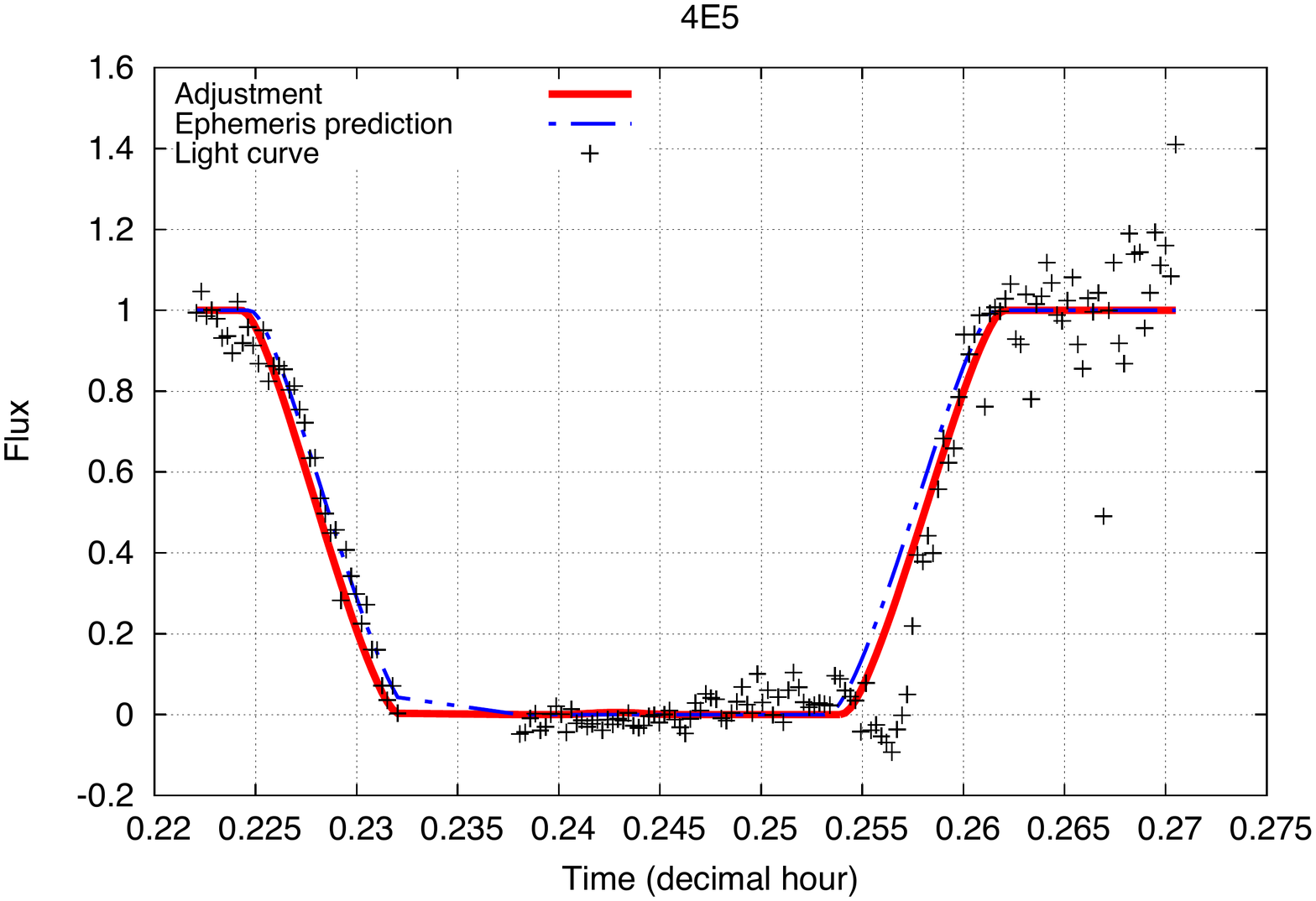}}
\caption{{\normalsize Fitted light curve (flux versus time in decimal hour) of the 4E5 event. Callisto eclipses Amalthea on 2015 Jan 7. The blue dot-dash line denotes the theoretical flux drop according to JUP310 ephemeris \citepads{JUP310}; the red line denotes the fitted flux drop of the event.}}
\label{Fig4E5}
\end{figure}

\begin{figure}
\centering
\resizebox{\hsize}{!}{\includegraphics{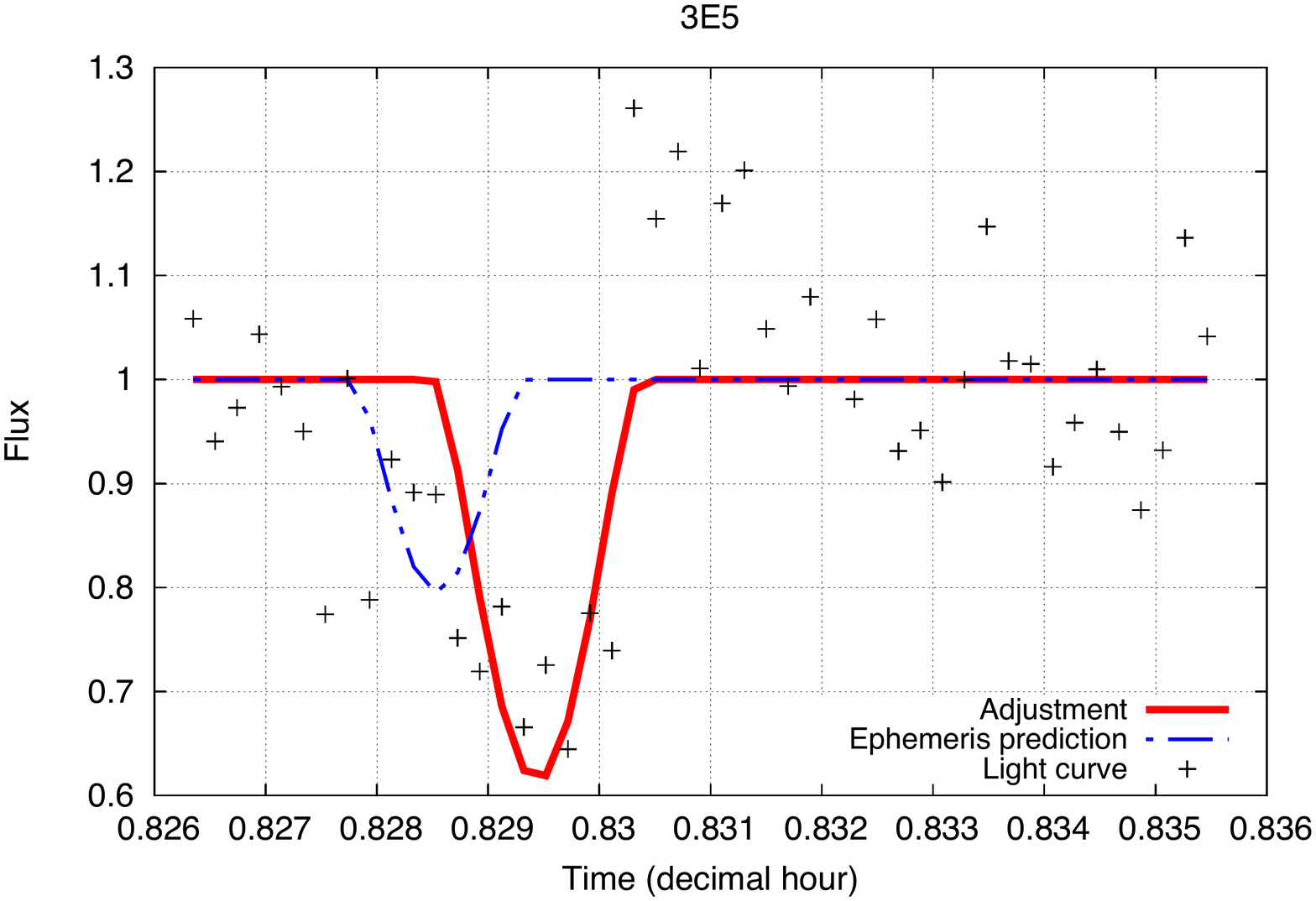}}
\caption{ {\normalsize Fitted light curve (flux versus time in decimal hour) of the 3E5 event. Ganymede eclipses Amalthea on 2015 Apr 7. The blue dot-dash line denotes the theoretical flux drop according to JUP310 ephemeris \citepads{JUP310}; the red line denotes the fitted flux drop of the event.}}
 \label{Fig3E5}
 \end{figure}

We did not use a real focal coronagraph, but a small suitable mask placed over the primary on the filter. As a consequence, an important amount of light remains visible on the image and is due to the diffusion in the different mirrors of the telescope  (Fig. \ref{FigPic}). Before extracting the photometric data, we had to remove this significant background that could affect the measurements. The idea was to isolate the satellite in a square of 20 by 20 pixels. Then, we removed this square and interpolated the artificial gap to rebuild the background that we fitted with a three- or five-degree polynomial. We finally removed the background polynomial \citepads{1991A&A...252..402C}. The light curves according to this reduction are provided in Figs. \ref{Fig4E6}, \ref{Fig4E5}, and \ref{Fig3E5} with crosses.

\begin{figure}
\centering
\resizebox{\hsize}{!}{\includegraphics{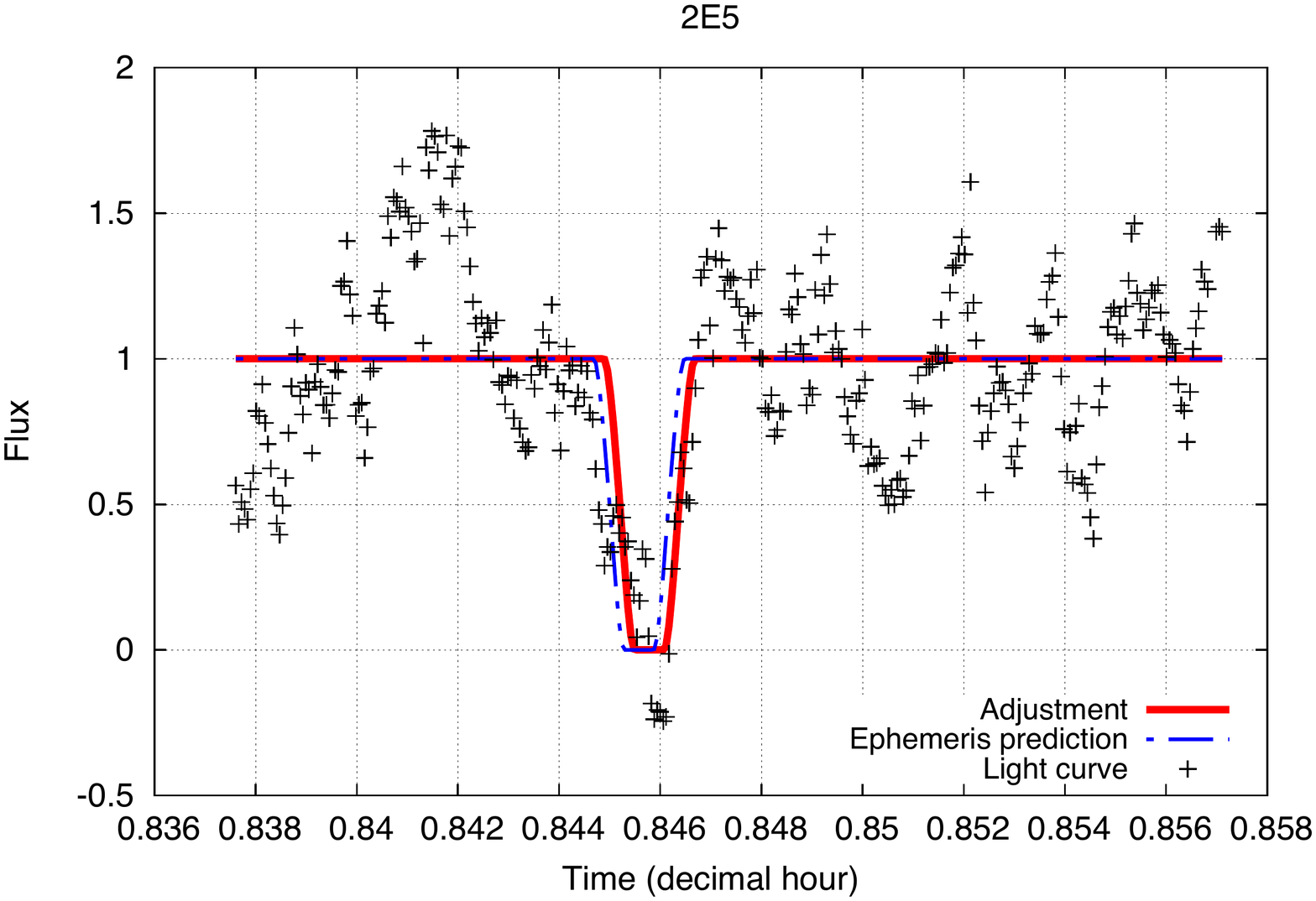}}
\caption{{\normalsize Fitted light curve (flux versus time in decimal hour) of the 2E5 event. Europa eclipses Amalthea on 2015 Apr 8. The blue dot-dash line denotes the theoretical flux drop according to JUP310 ephemeris \citepads{JUP310}; the red line denotes the fitted flux drop of the event.}}
\label{Fig2E5}
\end{figure}

\subsection{Observation at Saint-Sulpice Observatory}

The 2E5 observation on 2015 Apr 8 was made by two amateur observers: B. Christophe and O. Dechambre, at Saint-Sulpice Observatory, France (IAU Observatory Code: 947). They used a 60 cm telescope with a EMCCD Camera Raptor Photonics with a 890 nm Edmund filter to decrease the intensity of the planet and recorded 360 frames for the event.

We had to stack the images owing to the low signal of Amalthea to measure the photometric light flux during the event. A sliding average with 23 frames was realized. Then, the same method as for Pic du Midi observations was used to remove the background and determine the photometry of the satellite. The light curve according to this reduction is provided in Fig. \ref{Fig2E5} with crosses.


\section{Astrometric reduction and results}

\subsection{Astrometric reduction}

Photometric observations of the eclipses of the inner satellites of Jupiter consist in measuring the satellite flux during the events. Extracting astrometric data from such photometry is an important but difficult task. We propose a photometric model that is convenient for our purpose.\\

\noindent The light flux of the satellite $E(t)$ is measured at the time $t_1$, denoted as $K=E(t_1)$. During the event, the flux can be expressed as
\begin{equation*}
E(t)=K\cdotp S(t),
\end{equation*}
where $S(t)<1$ during the eclipse, and $S(t)=1$ before and after the event. With the function $S(t)$ defined in such a way, we have to use more realistic relations,
\begin{equation*}
E(t)=\left[K+K_t(t-t_0)\right]\cdotp S(t) + \left[P+P_t(t-t_0)\right],
\end{equation*}
where $t_0$ is an arbitrary fixed time and the constants $K$, $K_t$, $P$, $P_t$ have to be found from the photometric observation of the event. We then consider an axis of the event from the center of the Sun to the center of the eclipsing satellite. We consider a heliocentric celestial reference frame made identical with the International Celestial Reference Frame (ICRF) by ephemerides used to implement the method. We  consider right ascension $\alpha$ and declination $\delta$ of the satellites. The function $S(t)$ depends on the angular heliocentric distance $d$ of the eclipsed satellite center from the axis of the event, and the distance $d$ depends, in turn, on the angular heliocentric geoequatorial coordinates $(X,Y)$ of the eclipsing satellite, with respect to the eclipsed one, i.e.,  
\begin{equation*}
X=(\alpha_a-\alpha_p)\cos\delta_p,\\
Y=\delta_a-\delta_p,
\end{equation*}
where $\alpha_a$, $\delta_a$ are heliocentric geoequatorial coordinates of the eclipsing (active) satellite calculated at the time $t_1$, and $\alpha_p$ and $\delta_p$ are heliocentric geoequatorial coordinates of the eclipsed (passive) satellite calculated at the time $t_2$, whereas $t_2-t_1$ is the time of light propagation between the eclipsing and eclipsed satellites, and  $t_2-t_{obs}$ is the time of light traveling from the eclipsed satellite being observed at $t_{obs}$. The $(X, Y)$ coordinates are the so-called pseudoheliocentric coordinates, inherent to mutual eclipse. With sufficient precision, the angle $d$ can be calculated:
\begin{equation*}
d=\sqrt{X^2+Y^2}.
\end{equation*}

\noindent We adopt the following simplifications: the homogeneous disk of the eclipsed satellite is seen from the Earth just as from the Sun, and the eclipsing satellite and the Sun are spherical and the eclipsed satellite moves in the plane of the event, being perpendicular to the axis of the event. With our method, we calculate an elementary flux as one coming from an arc on the eclipsed satellite's disk, which is centered on the intersection of the event plane with the axis of the event. We then integrate the flux for all such arcs. In consequence, the following relation can be used for our purposes:
\begin{equation*}
E(t)=\left[K+K_t(t-t_0)\right]\cdotp S(X,Y) + \left[P+P_t(t-t_0)\right].
\end{equation*}
The true values of the $(X,Y)$ coordinates can be presented as 
\begin{equation*}
X= X_{th}(t)+D_x,\\
Y=Y_{th}(t)+D_y,
\end{equation*}
where $(X_{th}(t), Y_{th}(t))$ are the coordinates calculated from the ephemerides, and  $D_x$ and $D_y$ are constant parameters for the event. If we consider a single measurement at time $t_i$, the equations can be written as

\begin{multline}
\label{condequat}
E(t_i)=\left[K+K_t(t_i-t_0)\right]\cdotp S(X_{th}(t_i)+D_x,Y_{th}(t_i)+D_y) + \\
\left[P+P_t(t_i-t_0)\right].
\end{multline}
We linearize conditional equations with respect to the $D_x$ and $D_y$ parameters. The $(X_{th}(t_i),Y_{th}(t_i))$ pseudoheliocentric coordinates for the time of observation $t_i$ are calculated with the MULTI-SAT natural satellite ephemeris server \citepads{2008A&A...487..759E}.

\begin{figure}
\centering
\includegraphics[width=\hsize]{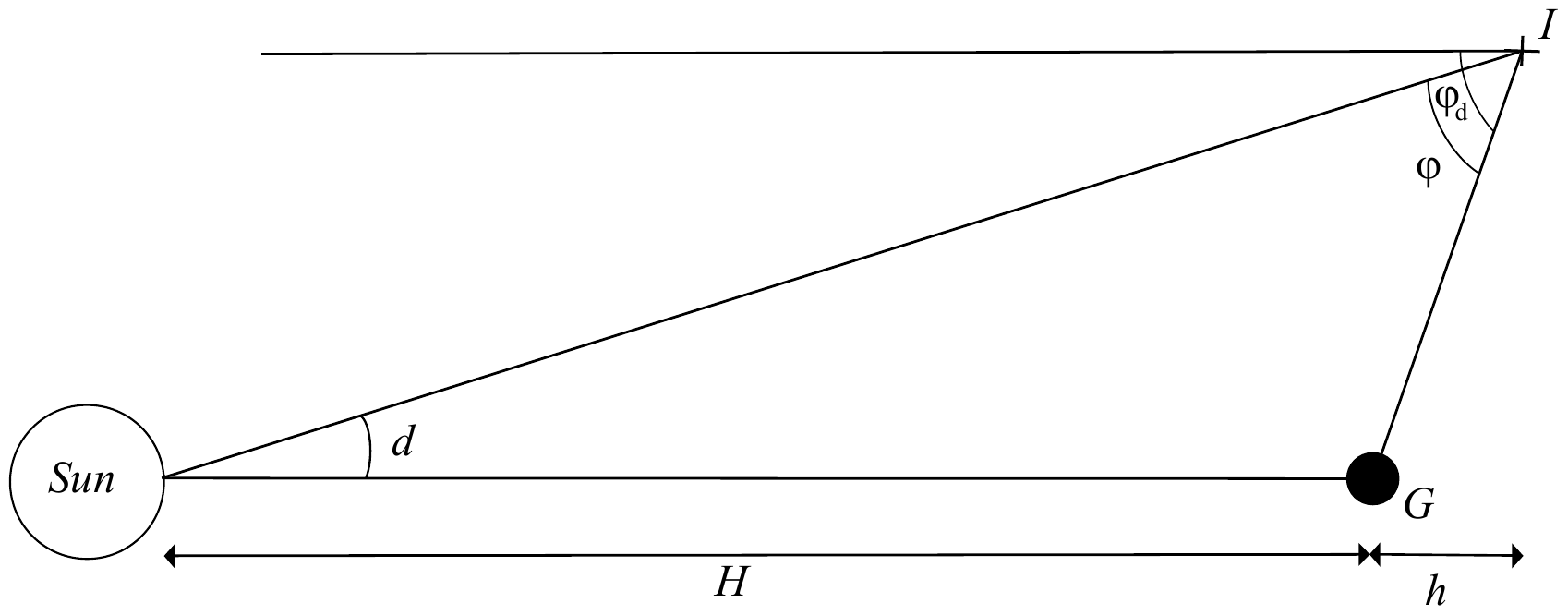}
\caption{{\normalsize Configuration representing the Sun, a Galilean satellite (G), an inner moon (I), and the different angles and distances in question.}}
\label{Schema}
\end{figure}

\noindent Since the photometric measurements are realized at times $t_i \; (i=1, 2, 3, ..., N)$, we consider a system with $N$ conditional equations (\ref{condequat}), with respect to the unknown parameters $D_x$, $D_y$, $K$, $K_t$, $P$, and $P_t$. An estimation of these parameters from observations can be determined by least squares.  Finally, we have to calculate the normalized $S(d)$ flux from a surface point of eclipsed satellite for a given angular heliocentric distance $d$ between this point and the center of eclipsing satellite. This point is illuminated by a part of a Sun disk that is not occulted by the eclipsed satellite. If the Sun disk is not occulted at all, it means that $S(d)=1$. Actually, $S(d)$ is determined as a non-occulted share of the Sun disk, and it depends in turn on the angular distance $\varphi$ between the centers of the Sun and the eclipsing satellite as it is seen from the point in question. To find $\varphi$ as a function of $d$ and $S(d)$, let us introduce the following notations (Fig. \ref{Schema}): 
\begin{itemize}
\item $H$ is the distance between the Sun and the eclipsing satellite;
\item $h$ is the distance between the eclipsed and eclipsing satellites;
\item $R$ is the radius of the Sun;
\item $r$ is the radius of the eclipsing satellite. 
\end{itemize}
So $\varphi$ can be determined using the following equations:

\begin{equation*}
\tan \varphi_d=\frac{H+h}{h} \tan d,\\
\varphi=\varphi_d-d.
\end{equation*}
Then $S(d)$ is determined with the sequence of formulas:
\begin{equation*}
r_1=\frac{r}{h},\\
r_2=\frac{R}{H+h}
\end{equation*}
\begin{equation*}
b=\sqrt{2\varphi^2(r_1^2+r_2^2)-\varphi^4-(r_1^2-r_2^2)^2 }
\end{equation*}
\begin{equation*}
a_1=\varphi^2-r_2^2+r_1^2,\\
a_2=\varphi^2-r_1^2+r_2^2
\end{equation*}
\begin{equation*}
\tan \varphi_1=\frac{b}{a_1},\\
\tan \varphi_2=\frac{b}{a_2}
\end{equation*}
\begin{equation*}
S(d)=1-\frac{1}{\pi}\left[\frac{r_1^2}{r_2^2}\varphi_1 +\varphi_2-\frac{b}{2r_2^2}\right]
\end{equation*}
In the case where $\varphi>r_1+r_2$, we note that we have $S(d)=1$. If $\varphi<|r_1-r_2|$, the result depends on the relation between $r_1$ and $r_2$. If $r_2>r_1$, we have $S(d)=(r_2^2-r_1^2)/r_2^2$  else $S(d)=0$.  Here we followed the formulas and conclusions given in  \citetads{1995ARep...39..539E}. In this way, the $D_x$ and $D_y$ parameters are determined, and the astrometric result is expressed as the pseudoheliocentric $(X,Y)$ coordinates at some time $t^*$ inside the observed event calculated by the relations  
\begin{equation}
X= X_{th}(t^*)+D_x,\\
Y=Y_{th}(t^*)+D_y
\label{finequat}
.\end{equation}

Of course there are different situations. If the measured flux before and after the event is approximately the same, the parameters $K_t$ and $P_t$ can be equal to zero. Otherwise, and in the case of partial mutual eclipse, $P$ cannot  be determined from observations and consequently the  only possibility is to use $P=0$. Finally, and in the case of a full or annular mutual eclipse, $D_y$ also cannot be determined from observations. We then use the following procedure with new $A$ and $B$ associated parameters
\begin{equation*}
D_x= A \cos T + B \sin T,\\
D_y=-A \sin T + B \cos T,
\end{equation*}
where $T$ is a tilt of the apparent track of the eclipsing satellite with respect to the eclipsed one. The $T$ tilt is the angle between the apparent relative velocity of the eclipsing satellite and the $X$-axis measured counterclockwise. The value of $T$ can be found from the $(X_{th}(t_i),Y_{th}(t_i))$ ephemeris coordinates. We fix B given by the ephemeris and determine $A$ from observations with other parameters. After we find the parameters, we search for time $t^*$ of minimal apparent distance between the satellites. The astrometric result is the heliocentric position angle of the eclipsing satellite relative to the eclipsed one at the time $t^*$, and the position angle is equal to the tilt $T$.

\subsection{Results and discussions}

\begin{figure}
\centering
\resizebox{\hsize}{!}{\includegraphics{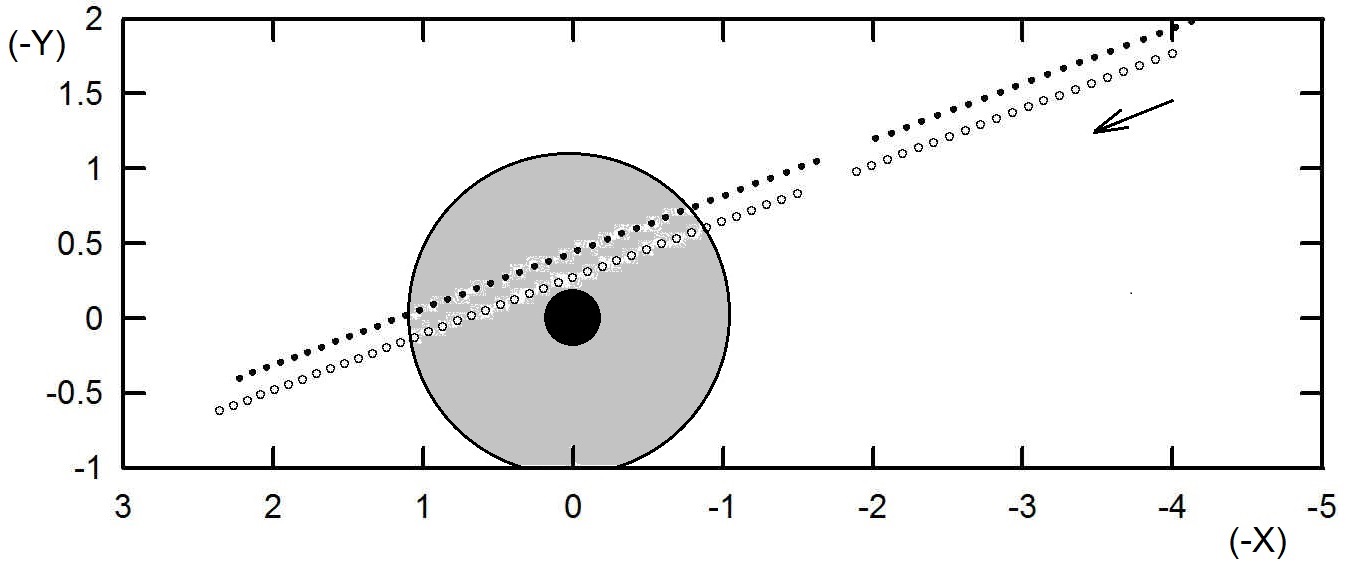}}
\caption{{\normalsize Apparent track of Thebe during the phenomena 4E14 on 2015 Jan 7 relative to the shadow and penumbra of Callisto before the fit of the parameters (circles) and after the fit (points). Units are in arcseconds.}}
\label{track_4E6}
\end{figure}

\begin{figure}
\centering
\resizebox{\hsize}{!}{\includegraphics{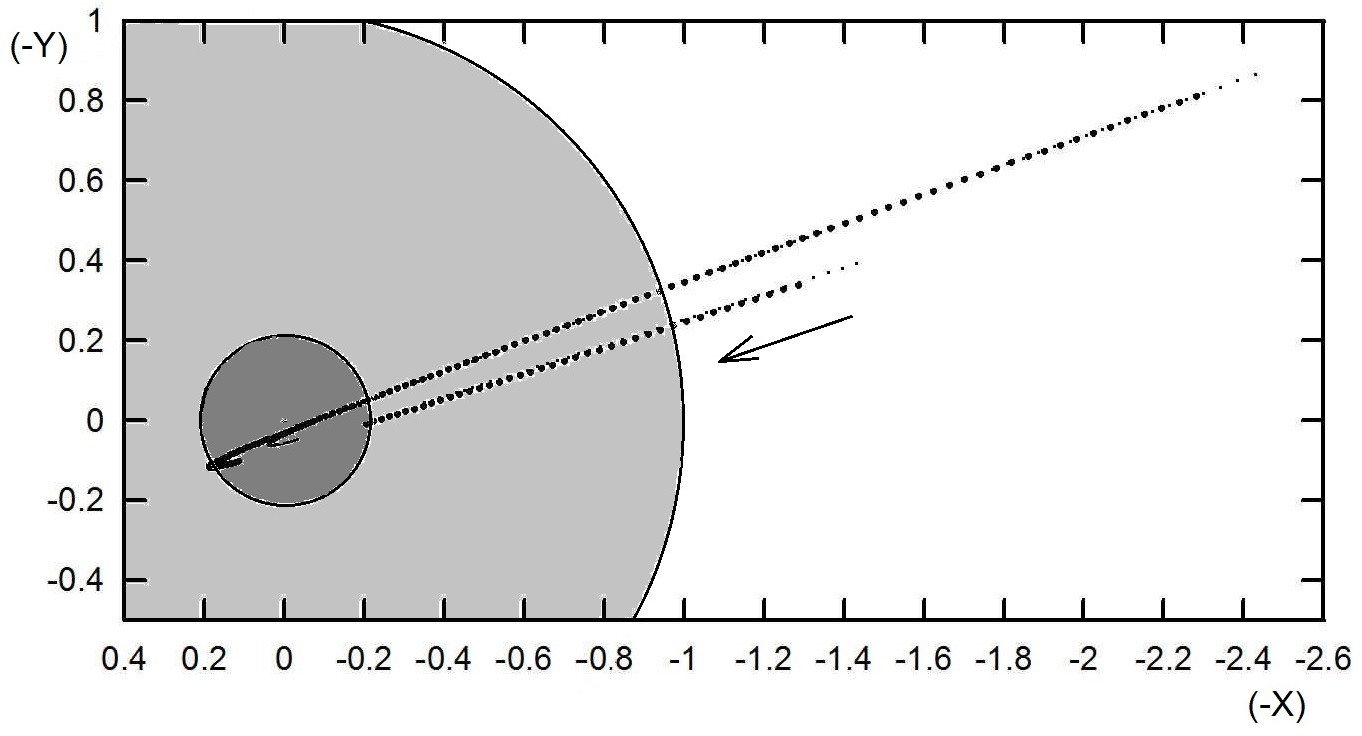}}
\caption{{\normalsize Apparent track of Amalthea during the phenomena 4E5 on 2015 Jan 7 relative to the shadow and penumbra of Callisto before the fit of the parameters (small dots) and after the fit (large dots). The shift is along the track and the small dots are only visible in a small area at the right. Units are in arcseconds.}}
\label{track_4E5}
\end{figure}

\begin{figure}
\centering
\resizebox{\hsize}{!}{\includegraphics{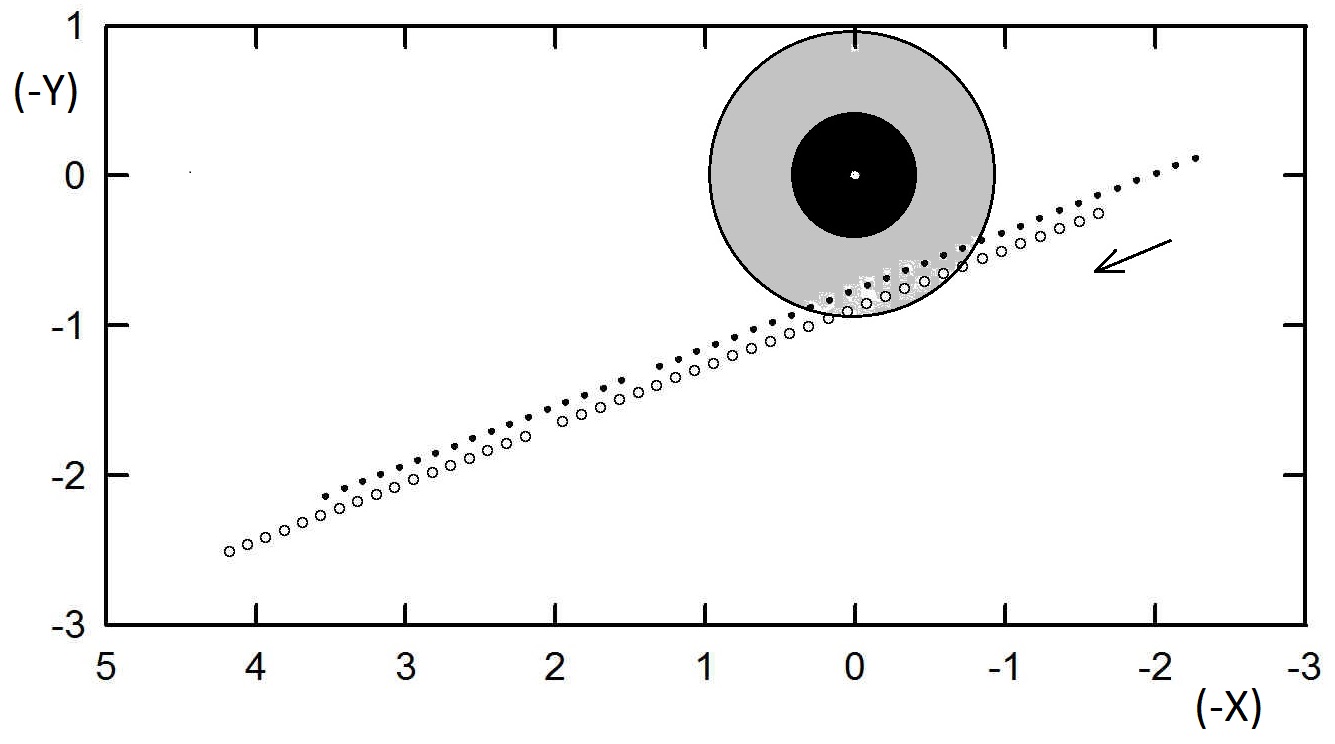}}
\caption{{\normalsize Apparent track of Amalthea during the phenomena 3E5 on 2015 Apr 7 relative to the shadow and penumbra of Ganymede before the fit of the parameters (circles) and after the fit (points). Units are in arcseconds.}}
\label{track_3E5}
\end{figure}

\begin{figure}
\centering
\resizebox{\hsize}{!}{\includegraphics{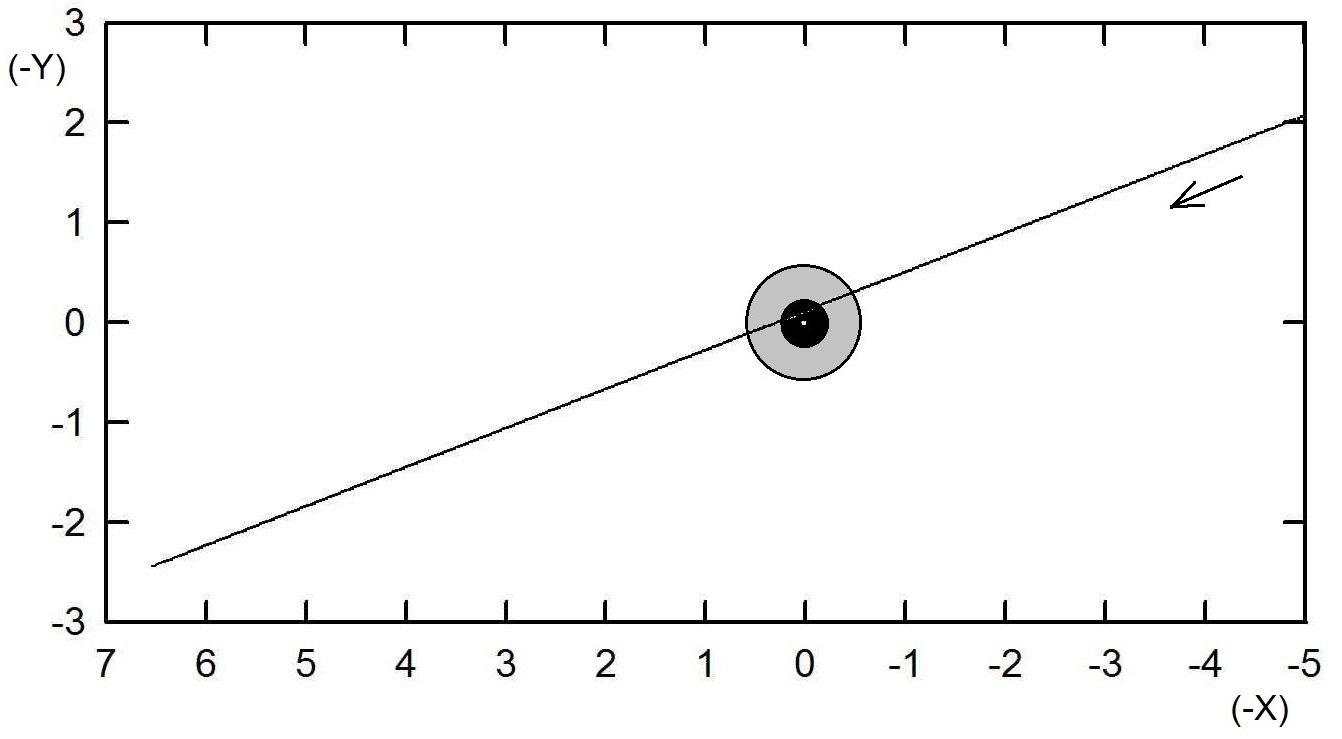}}
\caption{{\normalsize Apparent track of Amalthea during the phenomena 2E5 on 2015 Apr 8  relative to the shadow and penumbra of Europa before the fit of the parameters (small dots) and after the fit (large dots). Units are in arcseconds.}}
\label{track_2E5}
\end{figure}

Results of the astrometric reductions according to JUP310 are provided in Table \ref{table:3}. Because this model cannot be directly used for pseudoheliocentric coordinates to introduce in equation \ref{finequat}, we constructed a precessing model from its theoretical data. We also provide uncertainties on the coordinates at 1$\sigma$ given by least squares that are mainly due to random error in the photometry. For the observations of full eclipse for which we could not determine pseudoheliocentric coordinates, we provide positioning accuracies along the relative apparent path with the astrometric errors, at $1\sigma$, under the column ``Sigma along the orbit''. The apparent tracks of the eclipsed satellites are shown in Figs. \ref{track_4E6}-\ref{track_2E5} as they would be seen from the Sun. We present in these figures the opposite coordinates ($-X$ and $-Y$) of the eclipsed satellite relative to the eclipsing one.

We also determined astrometric results with the \citetads{2008SoSyR..42..296A} model to look for variances on the measured $(X,Y)$ coordinates if different $(X_{th},Y_{th})$ data were used in equations \ref{finequat}. We found that the results were quite similar and conclude that the fits of the $D_x$ and $D_y$ parameters minimize the differences between the different sets of $(X_{th},Y_{th})$ coordinates used.

\begin{table*}
\caption{\normalsize{Results of the astrometric reduction for the observed mutual events, according to JUP310.}}             
\label{table:3}      
\centering                          
\begin{tabular}{c c c c c c c c c}        
\hline\hline                 
Date &Type & $t^*$         & X  & Y & Position  & Sigma\\  
        &         &      &  & & Angle & along the orbit\\
        &         & (hh:mm:ss) & (arcsec)&(arcsec)& (deg) & (arcsec)\\
\hline                        
\textbf{2015 Jan 7}&   4E14 & 04:50:07.220 & $-0.164 \pm 0.053$& $-0.377 \pm 0.045$ &  \\
\textbf{2015 Jan 7} &  4E5 & 05:38:27.028 & & & 20.452 & 0.001\\
\textbf{2015 Apr 7} &  3E5 & 19:54:30.495 & $0.206 \pm 0.062$ &  $0.682 \pm 0.023$& \\
\textbf{2015 Apr 8} &  2E5 & 20:17:56.440 &  &   & 201.413 & 0.023\\
\hline                                   
\end{tabular}
\end{table*}

Once the $X$ and $Y$ coordinates are obtained as astrometric results of the photometric observations, we can compare them with different ephemerides. For this purpose, we used the server MULTI-SAT described in \citetads{2008A&A...487..759E}. This server provides $(X,Y)$ pseudoheliocentric coordinates that are useful for mutual eclipse observations. In this way we obtained the differences between the reduced $(X,Y)$ coordinates and those computed with the Avdyushev \& Ban’shikova model, or those computed with JUP310. Differences are given in Table \ref{table:4} under the columns ``$(O-C)_{X/Av}$'', ``$(O-C)_{Y/Av}$'', ``$(O-C)_{X/JUP}$'', and ``$(O-C)_{JUP}$''. For full eclipse, as we did for 4E5 on 2015 Jan 7 and 2E5 on 2015 Apr 8, we computed  the corrections of the apparent relative position of the satellites along their orbit with our own software. We also computed the differences between the positions obtained with astrometric reduction and those given by the ephemerides. The differences are provided in Table \ref{table:4} under the column ``$(O-C)_{Av}$ along the orbit'' and ``$(O-C)_{JUP}$ along the orbit'' for both of the  models used.

\begin{table*}
\caption{\normalsize{(O-C) at $t^*$ according to Avduyshev (Av) and JUP310 (JUP).}}             
\label{table:4}      
\centering                          
\begin{tabular}{c c c c c c c c}        
\hline\hline                 
Date &Type & $(O-C)_{X/Av}$ & $(O-C)_{Y/Av}$ & $(O-C)_{X/JUP }$ & $(O-C)_{Y/JUP }$ & $(O-C)_{Av}$ along & $(O-C)_{JUP }$ along \\  
        &         &   &  &  && the orbit &the orbit \\
        &         & (arcsec)&(arcsec)& (arcsec)&(arcsec)& (arcsec)& (arcsec)\\
\hline                        
\textbf{2015 Jan 7}&   4E14 &   0.129 &  -0.215 & -0.073 & -0.120 & & \\
\textbf{2015 Jan 7} &  4E5 & & & & & -0.156 & -0.095 \\
\textbf{2015 Apr 7} &  3E5 & 0.643 &  -0.370 & 0.516 & -0.299 & & \\
\textbf{2015 Apr 8} &  2E5 & & &  &  & 0.131 &0.022\\
\hline                                   
\end{tabular}
\end{table*}

\subsubsection{Thebe}

We provide results for the record of the eclipse of Thebe. The precision of the astrometric results is about 50 mas along the $X$-axis and 45 mas along the $Y$-axis, or 150 km and 135 km at the distance of Jupiter, respectively. We measure a difference of 73 mas with JUP310 ($\sim 220$ km) along the $X$-axis and 120 mas ($\sim 360$ km) along the $Y$-axis. We conclude that the error on the astrometric result is smaller than that given by the model.

Since we only have this observation for Thebe, it is difficult to draw any conclusions on the reliability of its ephemeris, and it would be interesting to compare our data with direct astrometric observations. The last paper dealing with such work was written by \citetads{2008P&SS...56.1804K}. However, the observations were compared with the JUP203 ephemerides \citepads{Jacobson_1994} and are  different from the one used in this paper. We computed their $(O-C)$ with JUP310 and found differences for the observations of 2002 ($-350$ mas in $\alpha$ and $-1250$ mas in $\delta$).

\subsubsection{Amalthea}

The three events we recorded for Amalthea allow to us make some statements about its ephemerides. First of all, we  note that the $(O-C)$ are smaller for JUP310 than for Avdyushev \& Ban’shikova ephemeris.

For the long event 4E5 on 2015 Jan 7, there is a great difference between the theory and our astrometric results. We observe a difference of  -95 mas along the apparent trajectory of the satellite. Furthermore, our accuracy is much smaller since we achieved 1 mas. This difference could be explained by an error due to the dynamical model. Indeed, this phenomena occurred during Amalthea’s elongation. An error on the inclination of the orbit could produce such an error in the position angle prediction. 

For the eclipse 3E5 on 2015 Apr 7, we observe a large $(O-C)$. However, the conditions of this observation were extremely difficult and we were  not able to record the event as early as we wished because Amalthea had just emerged from  the mask. We also note some noise in the light curve in Fig. \ref{Fig3E5}. We computed a standard deviation of 0.11. The flux drop of this light curve is around 0.4,  which means that the level of noise is acceptable.

For the last event 2E5 on 2015 Apr 8, we observe a $(O-C)$ of 22 mas, corresponding to the uncertainties on the astrometric results. In  Fig. \ref{Fig2E5} we observe a significant level of noise. However, the standard deviation of the data is around 0.30, which is smaller than the flux drop of the light curve. 

\section{Conclusions}

We have presented and analyzed the observations of four mutual eclipses of the inner moons of Jupiter, Amalthea and Thebe, with the Galilean moons. In spite of the difficulties for ground-based observations of such faint objects located near the bright planet, we managed to obtain the  data we needed to build the light curves. 

We had to use a specific background treatment in order to make the photometry of the observations. This allowed us to calculate the astrometric position at  time $t^*$ with a precision of $53$ mas for the eclipse of Thebe, and to make comparisons with two models. The JUP310 ephemeris gave us a smaller $(O-C)$. The difference with our observations for Thebe is $120$ mas with an error of 45 mas, and the maximum difference for Amalthea is $-95$ mas with an error of 1 mas.

Observational data of these eclipses and the relative coordinates of the satellites obtained with the astrometric reduction are useful for quantifying their reliability for orbit determination work and also to improve the ephemerides.

We plan to make direct astrometry of these inner satellites to obtain more data and to compare the accuracy of both methods.

\begin{acknowledgements}
This work was partly supported by Campus France, by the Scientific Council of Paris Observatory, by IMCCE, and by IPSA. This work was supported by the Russian Foundation for Basic Research, project no. 16-52-150005-CNRS-a.
\end{acknowledgements}

\bibliographystyle{aa} 
\bibliography{biblio.bib} 

\begin{thebibliography}{13}
\expandafter\ifx\csname natexlab\endcsname\relax\def\natexlab#1{#1}\fi

\bibitem[{{Arlot} {et~al.}(2014){Arlot}, {Emelyanov}, {Varfolomeev},
  {Amoss{\'e}}, {Arena}, {Assafin}, {Barbieri}, {Bolzoni}, {Bragas-Ribas},
  {Camargo}, {Casarramona}, {Casas}, {Christou}, {Colas}, {Collard}, {Combe},
  {Constantinescu}, {Dangl}, {De Cat}, {Degenhardt}, {Delcroix},
  {Dias-Oliveira}, {Dourneau}, {Douvris}, {Druon}, {Ellington}, {Estraviz},
  {Farissier}, {Farmakopoulos}, {Garlitz}, {Gault}, {George}, {Gorda},
  {Grismore}, {Guo}, {Herald}, {Ida}, {Ishida}, {Ivanov}, {Klemt}, {Koshkin},
  {Le Campion}, {Liakos}, {Liao}, {Li}, {Loader}, {Lopresti}, {Lo Savio},
  {Marchini}, {Marino}, {Masi}, {Massall{\'e}}, {Maulella}, {McFarland},
  {Miyashita}, {Napoli}, {Noyelles}, {Pauwels}, {Pavlov}, {Peng},
  {Perell{\'o}}, {Priban}, {Prost}, {Razemon}, {Rousselle}, {Rovira}, {Ruisi},
  {Ruocco}, {Salvaggio}, {Sbarufatti}, {Shakun}, {Scheck}, {Sciuto}, {da Silva
  Neto}, {Sinyaeva}, {Sofia}, {Sonka}, {Talbot}, {Tang}, {Tejfel}, {Thuillot},
  {Tigani}, {Timerson}, {Tontodonati}, {Tsamis}, {Unwin}, {Venable},
  {Vieira-Martins}, {Vilar}, {Vingerhoets}, {Watanabe}, {Yin}, {Yu}, \&
  {Zambelli}}]{2014A&A...572A.120A}
{Arlot}, J.-E., {Emelyanov}, N., {Varfolomeev}, M.~I., {et~al.} 2014, \aap,
  572, A120

\bibitem[{{Arlot} \& {Stavinschi}(2007)}]{2007ASPC..370...58A}
{Arlot}, J.-E. \& {Stavinschi}, M. 2007, in Astronomical Society of the Pacific
  Conference Series, Vol. 370, Solar and Stellar Physics Through Eclipses, ed.
  O.~{Demircan}, S.~O. {Selam}, \& B.~{Albayrak}, 58

\bibitem[{{Avdyushev} \& {Ban'shikova}(2008)}]{2008SoSyR..42..296A}
{Avdyushev}, V.~A. \& {Ban'shikova}, M.~A. 2008, Solar System Research, 42, 296

\bibitem[{{Barnard}(1892)}]{1892AJ.....12...81B}
{Barnard}, E.~E. 1892, \aj, 12, 81

\bibitem[{{Burns} {et~al.}(1999){Burns}, {Showalter}, {Hamilton}, {Nicholson},
  {de Pater}, {Ockert-Bell}, \& {Thomas}}]{1999Sci...284.1146B}
{Burns}, J.~A., {Showalter}, M.~R., {Hamilton}, D.~P., {et~al.} 1999, Science,
  284, 1146

\bibitem[{{Christou} {et~al.}(2010){Christou}, {Lewis}, {Roche}, {Hidas}, \&
  {Brown}}]{2010A&A...522A...6C}
{Christou}, A.~A., {Lewis}, F., {Roche}, P., {Hidas}, M.~G., \& {Brown}, T.~M.
  2010, \aap, 522, A6

\bibitem[{Colas(1991)}]{colas_1991}
Colas, F. 1991, PhD thesis, Observatoire de Paris

\bibitem[{{Colas} \& {Arlot}(1991)}]{1991A&A...252..402C}
{Colas}, F. \& {Arlot}, J.~E. 1991, \aap, 252, 402

\bibitem[{{Emel'Yanov}(1995)}]{1995ARep...39..539E}
{Emel'Yanov}, N.~V. 1995, Astronomy Reports, 39, 539

\bibitem[{{Emel'Yanov} \& {Arlot}(2008)}]{2008A&A...487..759E}
{Emel'Yanov}, N.~V. \& {Arlot}, J.-E. 2008, \aap, 487, 759

\bibitem[{Jacobson(1994)}]{Jacobson_1994}
Jacobson, R. 1994, Revised ephemerides for the inner Jovian satellites.

\bibitem[{Jacobson(2013)}]{JUP310}
Jacobson, R. 2013, JUP 310 Release

\bibitem[{{Kulyk}(2008)}]{2008P&SS...56.1804K}
{Kulyk}, I. 2008, \planss, 56, 1804

\end{thebibliography}
\end{document}